# Analytical models of probability distribution and excess noise factor of Solid State Photomultiplier signals with crosstalk

S. Vinogradov

P.N. Lebedev Physical Institute of the Russian Academy of Sciences, Moscow, Russia

E-mail: vin@lebedev.ru

**Abstract**

Silicon Photomultipliers (SiPM), also so-called Solid State Photomultipliers (SSPM), are based on Geiger mode avalanche breakdown limited by strong negative feedback. SSPM can detect and resolve single photons due to high gain and ultra-low excess noise of avalanche multiplication in this mode. Crosstalk and afterpulsing processes associated with the high gain introduce specific excess noise and deteriorate photon number resolution of the SSPM. Probabilistic features of these processes are widely studied because of its high importance for the SSPM design, characterization, optimization and application, but the process modeling is mostly based on Monte Carlo simulations and numerical methods. In this study, crosstalk is considered to be a branching Poisson process, and analytical models of probability distribution and excess noise factor (ENF) of SSPM signals based on the Borel distribution as an advance on the geometric distribution models are presented and discussed. The models are found to be in a good agreement with the experimental probability distributions for dark counts and a few photon spectrums in a wide range of fired pixels number as well as with observed super-linear behavior of crosstalk ENF.

## 1. Introduction

Silicon or, more general, Solid State Photomultipliers (SiPM, SSPM) are widely recognized as photodetectors with the perfect pulse height (photon number) resolution due to high gain and ultra-low excess noise of a limited Geiger mode avalanche multiplication. However, high gain of the SSPM is inherently accompanied with hot-carrier-induced optical emission and trapping-detrapping effects resulting in crosstalk and afterpulsing, each of which in turn deteriorates the probability distribution of output signals and arises specific excess noises. Excess noise factor (ENF) of crosstalk and afterpulsing of SSPM without protection at high overvoltage happens to be much higher than that of vacuum photomultipliers (PMT). It means that actual resolution of SSPM may be much worse than that of PMT even if a few photon spectrums of SSPM output signals are clearly resolved [1-4].

Therefore, we have to understand the role of the crosstalk and afterpulsing probability distributions and corresponding excess noises in the SSPM design (e.g. trade-off in reduction of crosstalk by inter-pixel trench isolation and photon detection efficiency (PDE) by fill factor), characterization (e.g. pure single electron gain and single photon PDE), optimization (e.g. optimal overvoltage to balance PDE and crosstalk for lowest ENF), and application (e.g. optimal overvoltage and integration time for the best energy resolution of scintillation detection).

Obviously, analytical expressions for the probability distribution, and, hence, mean, variance and ENF of crosstalk and afterpulsing may provide complete knowledge about these stochastic processes. In spite of considerable interest to crosstalk and afterpulsing expressed almost in all studies of SSPM, the theoretical analyses of crosstalk and afterpulsing contributions to the SSPM output signals are mostly based on Monte Carlo simulations. This approach often provides good fit with experimental results (e.g. [4]), but it could not deliver analytical clarity in our understanding of these effects. Some studies are focused on recursive algorithms and linear matrix transformation approaches to reconstruct SSPM signals with crosstalk, but the higher-order cascade effects are still used to be accounted by numerical methods [5-7].

This study presents and discusses the analytical models of crosstalk as branching Poisson processes based on the Borel distribution. These new models represent an advance on the geometric chain process models based on the geometric distribution of crosstalk events reported earlier [8]. Both approaches are comparatively applied to some known experimental results of previously published studies of the SiPM with crosstalk.

## 2. Analytical models of probability distribution

### 2.1. General consideration

Let us consider a general case of SSPM output signal statistics. We assume that any output pulse produced by any single fired pixel – so-called single electron response (SER) – is identical regardless the pixel triggering origin (dark electron, photoelectron, crosstalk). Appearance of SER is further referenced as an event. Simultaneously fired pixels yield multiple SER events. We assume that crosstalk events – secondaries – appear simultaneously with the initially fired pixels – primaries – due to photoelectron or dark electron triggering.

Total number of output SER events initiated by $N$ primary events – random variable $X$ – includes all primaries and all secondaries as follows:

$$X = \sum_{i=1}^{N}(1+C_i), \qquad (1)$$

where $C_i$ – number of crosstalk events initiated by single primary event $i$. $C_i$ are assumed to be independent identically distributed random variables independent from $N$.

However, applicability of these assumptions is limited. Independence in co-existence of several simultaneous crosstalk chains may be more or less expressed in case if for any crosstalk event in any chain the number of neighboring pixels available for triggering does not depend on number of chains. This case corresponds to the hypothesis of considerable long-distance optical crosstalk originated from infrared part of hot-electron photon emission spectrum as discussed in [9-10], for example, through reflections from back surface of SiPM substrate. In any case, the number of chains and the total number of output events $X$ should be much less than number of the SiPM pixels.

Nevertheless, these assumptions allow to derive very simple analytical results, which may be useful at least as an upper margin estimation of crosstalk effects.

We are interested in common statistic properties of *X*, namely the probability distribution function *P(X=k)*, mean *E[X]*, and variance *Var[X]*. It should be noted that ENF of random output events has to be defined by two distinct ways depending on statistic nature of primary events.

If number of primary events *N* is a non-random (say, *N≡1*) as it happens in case if we measure SSPM output signal amplitude or charge histogram by triggering acquisition from 0.5 SER level (say, measuring dark counts) then *X* and *ENF(X)* are expressed as

$$X = 1 + C_1, \qquad (2)$$

$$ENF(X) = 1 + \frac{Var[X]}{E^2[X]}. \qquad (3)$$

This is well-known definition of ENF for, say, avalanche multiplication process, because in APD we consider gain (equal to *X*) resulting from single electron initiation of an avalanche.

Another approach should be used if *N* is a random number of primary events as it happens in photodetection. It means that the primaries themselves are a noisy input, which produces more noisy output due to contribution of noisy secondaries. Accordingly with this approach (well known for characterization of amplifiers), we have to define ENF as relative losses in signal to noise ratio (SNR) from input to output:

$$ENF = \frac{SNR^2(N)}{SNR^2(X)} = \frac{E^2[N]/Var[N]}{E^2[X]/Var[X]}. \qquad (4)$$

In the most common photodetection case, the number of photons and then the number of photoelectrons follow Poisson distribution, so *E[N]=Var[N]*. However, both approaches (3) and (4) yield the same result in correspondence with Burgess variance theorem applied for random amplification process with Poisson input as, for example, discussed in [11].

## 2.2. Geometric chain process

If one primary event produces one chain of secondary events with the same probability $p$ for each secondary event then total number of events $X$ obeys geometric distribution. This model is based on assumption that one preceding event may produce 0 or 1 succeeding event only (no double, triple etc. successors in one generation). This situation is shown in upper left scheme of Table 1 as a single chain consisting of the single primary event (black circle) then the first and the second crosstalk events (gray circles), and absence of the third crosstalk event (empty circle). Such model looks rather relevant for afterpulsing processes when the initially triggered pixel may be either retriggered or not, and every new triggering has to "erase" its previous trapping state. This is also applicable for crosstalk at low $p$, if we could consider probability of double events $\sim p^2$ as negligible. In photon detection case, if the number of primary events $N$ is a Poisson random variable with the mean $\mu$ then $X$ follows Compound Poisson distribution, as it was shown in [8] using generating function approach. Upper left scheme of Table 1 reflects this case as random number of geometric chains (1, 2, 3 …) with random crosstalk events in each (0, 1, 2…).

| Crosstalk Models | Single primary event $N\equiv1$ e.g. SSPM Dark Spectrum | Poisson number of primaries $<N>=\mu$ e.g. SSPM Photon Spectrum |
|---|---|---|
| Geometric Chain Crosstalk Process | 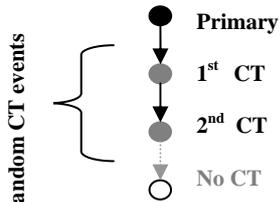 | 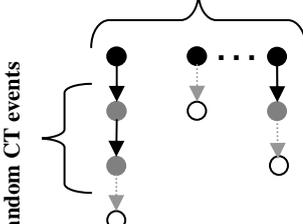 |
| Branching Poisson Crosstalk Process | 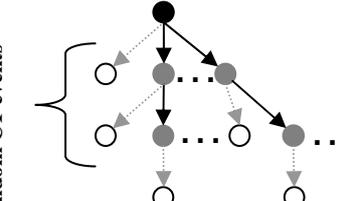 | 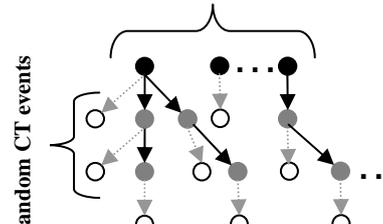 |

**Table 1.** Schematic overview of the crosstalk process models.

## 2.3. Branching Poisson process

However, it would be reasonable to consider crosstalk events as a result of a large number of trials to trigger neighboring pixels with small probability of success in each trial thus assuming Poisson distribution of the successful trials in a single (direct) generation of succeeding events. It means that one preceding event produces Poisson distributed random number of succeeding events and then again, until extinction, representing a branching Poisson process, as shown in lower left scheme of Table 1.

Probability distribution of $X$ in branching processes may be found using recursive generating function approach. In case of Poisson distributed number of single (direct) generation successors with the mean $\lambda$, the total number of events $X$ originated from single non-random primary event follows the Borel distribution.

Moreover, in case of Poisson distributed primary events $N$ with the mean $\mu$ the total number of all events $X$ is found to be a compound Poisson sum of Borel distributed random variables (lower right scheme of Table 1). Thus, the statistics of $X$ are governed by Generalized (Lagrangian) Poisson distribution [12].

Essential expressions of our interest for the both models are summarized in Table 2.

| Distribution | Geometric chain process | | Branching Poisson process | |
|---|---|---|---|---|
| Primary event distribution | Non-random single ($N\equiv 1$) | Poisson ($\mu$) | Non-random single ($N\equiv 1$) | Poisson ($\mu$) |
| Total event distribution | Geometric ($p$) | Compound Poisson ($\mu, p$) | Borel ($\lambda$) | Generalized Poisson ($\mu, \lambda$) |
| $P(X=k)$ | $p^{k-1} \cdot (1-p)$<br>$k = 1, 2\ldots$ | Ref. [8]<br>$k = 0, 1, 2\ldots$ | $\dfrac{(\lambda \cdot k)^{k-1} \cdot \exp(-k \cdot \lambda)}{k!}$<br>$k = 1, 2\ldots$ | $\dfrac{\mu \cdot (\mu + \lambda \cdot k)^{k-1} \cdot \exp(-\mu - k \cdot \lambda)}{k!}$<br>$k = 0, 1, 2\ldots$ |
| $E[X]$ | $\dfrac{1}{1-p}$ | $\dfrac{\mu}{1-p}$ | $\dfrac{1}{1-\lambda}$ | $\dfrac{\mu}{1-\lambda}$ |
| $Var[X]$ | $\dfrac{p}{(1-p)^2}$ | $\dfrac{\mu \cdot (1+p)}{(1-p)^2}$ | $\dfrac{\lambda}{(1-\lambda)^3}$ | $\dfrac{\mu}{(1-\lambda)^3}$ |
| ENF | $1+p$ | | $\dfrac{1}{1-\lambda} \sim 1 + p + \dfrac{3}{2}p^2 + o(p^3)$ | |

**Table 2.** Summary of the essential analytical expressions for the crosstalk affected SiPM signals.

## 3. Models and experiments

### 3.1. Dark-event-initiated distribution

The most direct way to compare the models with experimental results is a single-primary-event-initiated distribution analysis. It is often based on measurements of Dark Count Rate (DCR) vs. counter threshold level representing complementary cumulative distribution function of the crosstalk itself (if overlapping of dark counts is negligible). The highest achieved threshold level (likely known for today) in such experiments is reported in [1] for 5x5 mm$^2$ SiPM without crosstalk suppression at -61˚C, and this result is reproduced in Fig.1 (open circles).

In order to compare the model with the experiment, crosstalk probability $p$ is calculated from two points in experimental data accordingly widely used estimation (including [1]):

$$p = \frac{DCR(1.5SER)}{DCR(0.5SER)}. \qquad (5)$$

However, (5) is valid only for negligible probability of the primary event coincidence (low DCR). Fitting parameters $p$ and $\lambda$ for geometric and Borel distributions follow relations:

$$P(X>1) = 1 - P_{geom}(X=1) = 1 - P_{borel}(X=1) = p = 1 - \exp(-\lambda). \qquad (6)$$

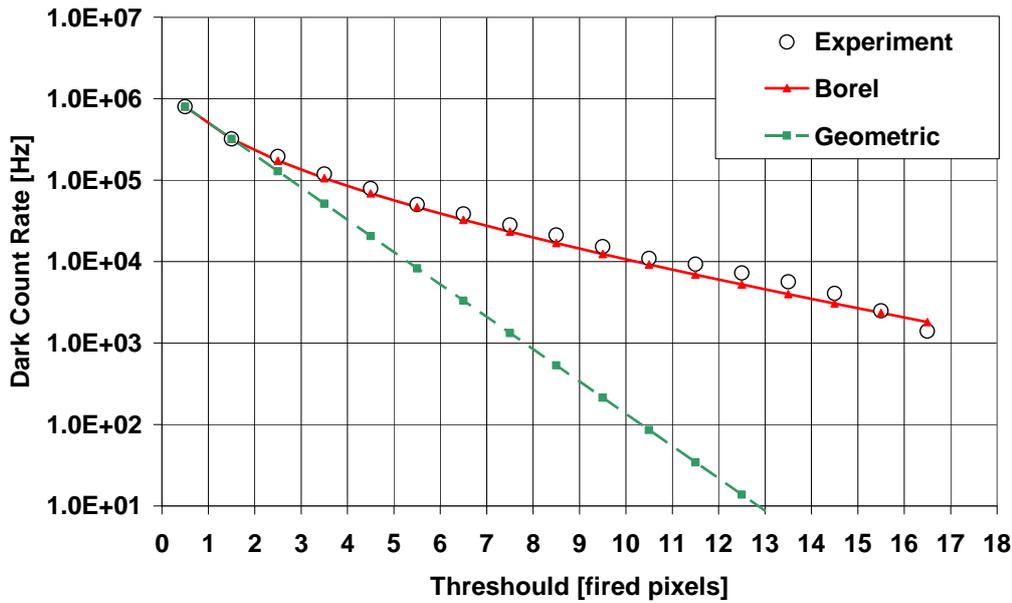

**Fig. 1.** DCR dependence on the counter threshold level in fired pixels (complementary cumulative distribution function of crosstalk events) measured for the 1600 pixels 5x5 mm$^2$ SiPM for MAGIC at -61˚C [1] in comparison with the analytical models.

Thus, crosstalk probability $p$ is found to be 40% and mean number of crosstalk events in a single generation $\lambda \sim 0.51$, then these values are used as a single fitting parameter for the model plots correspondingly. As shown, the Borel model fits well to the whole data set, but the geometric one cannot explain heavy tail at all.

### 3.2. Multi-photon-initiated distribution

The second way to compare the models with experimental results is based on a random-primary-event-initiated distribution measured in form of a short few photon pulse detection spectrum representing probability mass function of SSPM output signals. In this case, the crosstalk manifests itself as a deviation from pure Poisson law, i.e. as a supplementary effect. One of the most relevant example of such measurement (dynamic range of $10^4$, 11 fired pixels resolved) is presented in [6] for the 400 pixel MPPC (C10507-11-050U), and it is reproduced in Fig.2 (open circles). Both geometric chain and branching Poisson models are in a good agreement with the experiment because Borel distribution approaches to geometric one at low $p$ (~9%), and both of them coincide with the Poisson law at $p = 0$.

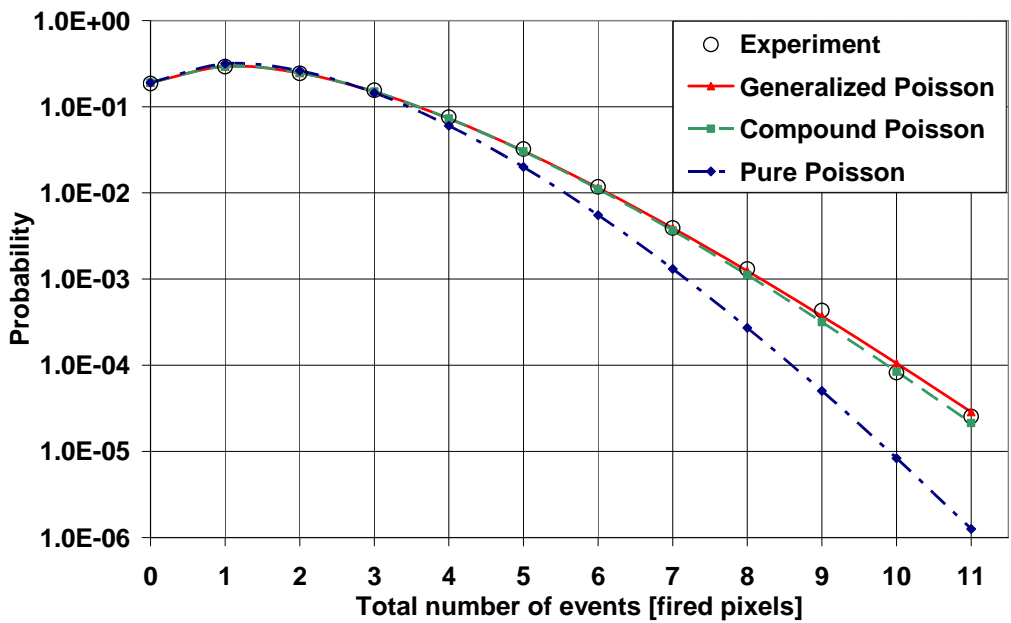

**Fig. 2.** A few photon detection spectrum at $\mu = 1.66$ (probability mass function of output signals with crosstalk) measured for the 400 pixels 1x1 mm$^2$ MPPC (C10507-11-050U) [6] in comparison with the analytical models.

### 3.3. Excess noise factor of multi-photon detection

Analysis of ENF is an indirect way of model verification, but it is very important for SSPM applications. Branching Poisson process model predicts less optimistic and more realistic behavior of ENF than geometric chain model, namely super-linear dependence *ENF(p)* providing better agreement with the experiments [1, 3, 4, 13].

An example of ENF dependence on crosstalk probability *p* is presented in Fig. 3, where Monte Carlo simulation results given in [4] are compared with the models using dependence of *p* on bias voltage from [4]. It should be mentioned that the Monte Carlo simulation approach in [4] assumes that contributions from excess noise of avalanche multiplication and dark noise are included in the resulting *ENF*. Obviously, as soon as the model ENF plots account for pure crosstalk contribution only, they appear to be lower than total ENF of SiPM under study at overvoltage < 2 V. However, SSPM saturation due to limited number of pixels would result in limited development of crosstalk process and thus lower *ENF* than predicted by the branching Poisson model. Particularly, it seems to be reflected in Fig. 3 as overestimated branching Poisson *ENF* plot at overvoltage > 2 V.

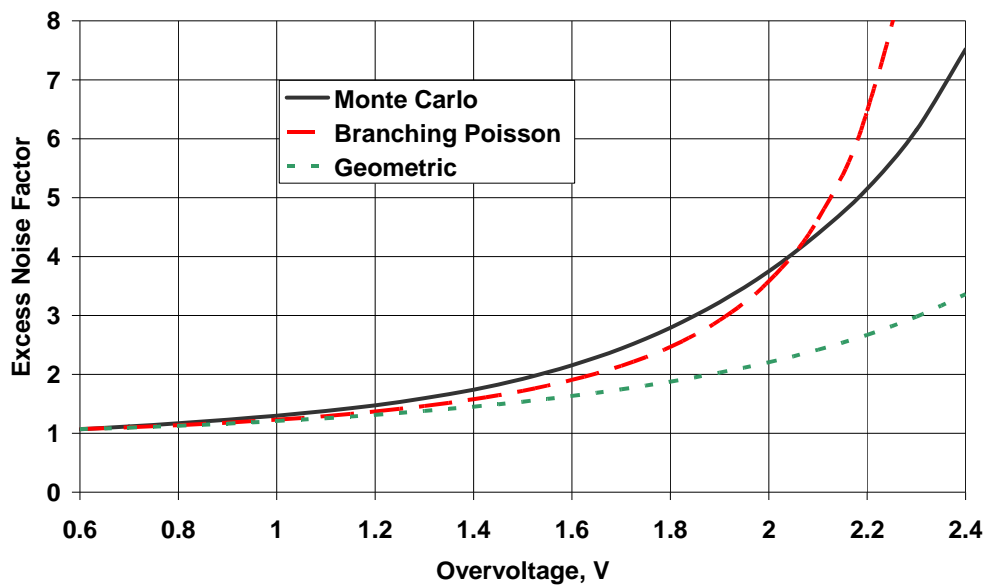

**Fig. 3.** Excess noise factor of SiPM simulated at 100 photons pulse detection in 540 ns gate for the 667 pixels 1.3x1.3 mm$^2$ MPPC for T2K [4] in comparison with the analytical models (Monte Carlo simulation also includes noise contributions resulting from avalanche multiplication and dark counts).

## 4. Conclusion

Branching Poisson process models provide simple analytical expressions for probability distributions and excess noise of SSPM signals with crosstalk. Good agreement of these expressions with at least a few well-known experimental results is found as preliminary verification of the models, but it has to be comprehensively studied further, especially in case of SSPM signals approaching to saturation. Saturation of SSPM is assumed to be one of the main reasons for variability – from very good fit to considerable deviation – in comparisons of the branching Poisson models with the experiments.

**References**

[1] P. Buzhan et al., Nucl. Instr. and Meth. A 610 (2009) 131.

[2] S. Vinogradov et al., IEEE Trans. Nucl. Sci. 58 (1) (2011) 9.

[3] Y. Musienko, in: SiPM Matching Event, CERN, Geneva, 16-17 Feb. 2011.

[4] A. Vacheret et al., arXiv:1101.1996, 2011.

[5] P. Eraerds, M. Legré, A. Rochas, H. Zbinden, N. Gisin, *Opt. Exp.* 15 (22) (2007) 14539.

[6] I. Afek et al., Phys. Rev. A 79 (2009) 7.

[7] M. Ramilli et al., J. Opt. Soc. Am. B 27 (2010) 852.

[8] S. Vinogradov et al., in: Proceedings IEEE Nuclear Science Symposium and Medical Imaging Conference (NSS/MIC 2009), Orlando, FL, 2009, 1496.

[9] R. Mirzoyan, R. Kosyra, H.-G. Moser, Nucl. Instr. and Meth. A 610 (2009) 98.

[10] A.N. Otte, Nucl. Instr. and Meth. A 610 (2009) 105.

[11] H.H. Barrett, in: H.H. Barrett, K.J. Myers, Foundation of Image Science, Wiley, Hoboken, N.J., 2004, 631-699.

[12] P.C. Consul, F. Famoye, Lagrangian Probability Distributions, Birkhäuser, Boston, 2006.

[13] S. Vinogradov et al., Bulletin of the Lebedev Physics Institute, submitted in 2010, in press.